\documentclass[aps,twocolumn,groupedaddress]{revtex4-1}
\usepackage{amsmath}
\usepackage{amssymb}
\usepackage{amsfonts}
\usepackage{graphicx}
\usepackage{dcolumn}
\usepackage{float}
\usepackage{bm}
\usepackage{mathrsfs}
\usepackage{txfonts}
\usepackage{tabularx}
\usepackage{bigstrut}
\usepackage{soul}
\usepackage[usenames,dvipsnames]{color}
\usepackage[hyperindex,pdftex]{hyperref}

\newcommand{\beq}{\begin{eqnarray}}
\newcommand{\eeq}{\end{eqnarray}}
\newcommand{\ua}{\uparrow}
\newcommand{\da}{\downarrow}
\newcommand{\dg}{\dagger}
\newcommand{\mbf}{\mathbf}
\newcommand{\la}{\langle}
\newcommand{\ra}{\rangle}
\newcommand{\bso}{\boldsymbol}

\newcommand{\intw}{\int^{\infty}_{-\infty}}

\begin{document}

\title{The optical properties of dibenzoterrylene}
\author{Z. S. Sadeq }
\email{sadeqz@physics.utoronto.ca}

\author{Rodrigo A. Muniz}

\author{J. E. Sipe}

\affiliation{Department of Physics, University of Toronto, Toronto, Ontario M5S 1A7, Canada}

\date{\today}
\begin{abstract}
Dibenzoterrylene (DBT) has garnered interest as a potential single photon source (SPS). To have a better grasp of any possible limitations of using DBT for this application, a better understanding of its optical properties is needed. We use a configuration interaction (CI) strategy to calculate the many body wavefunctions of DBT, and we use these wavefunctions to calculate its optical properties. We calculate the linear absorption spectrum and the spatial distributions of electrons involved in several bright transitions. We also calculate the two-photon absorption spectrum of DBT and show that there are several excited states that are bright due to two-photon absorption. Except at high photon energies, we predict that there are no competing optical processes regarding the use of DBT as a SPS. Our calculations provide details of the optical properties of DBT that are interesting in general, and useful for considering  optical applications of DBT. 

\end{abstract}
\maketitle

\section{Introduction}
Single photon sources (SPSs) are an important resource for optical based quantum information processing \cite{M.D.Eisaman2011,M.Schiavon2016,J.-B.Trebbia2009,B.Lounis2005,OBrien2007,C.Kurtsiefer2000,I.Aharonovich2011}. Candidate devices are based on semiconductor quantum dots \cite{M.Mueller2017,S.Buckley2012,P.Senellart2017,D.C.Unitt2005}, color centers in diamond \cite{C.Wang2006,I.A.Khramtsov2017,L.Marseglia2018}, and trapped atoms (or ions) in the gas phase \cite{D.B.Higginbottom2016,M.Hijlkema2007,M.Keller2004}. Organic materials at cryogenic temperatures also can act as a source of single photons; typically, the optical coherence lifetimes of the relevant transitions in organic materials are longer by an order of magnitude than those of semiconductor quantum dots \cite{J.-B.Trebbia2009, C.Toninelli2010, A.A.L.Nicolet2007,A-M.Boiron1996}. Synthesis of organic materials is relatively straightforward \cite{S.Faez2015, Y.Li2016, K.D.Major2015}; they are simple to deposit on optical chips and waveguides \cite{P.E.Lombardi2017}. Thus organic materials open up the possibility of using existing integrated chip strategies to carry out a variety of nonlinear optical processes \cite{C.Polisseni2016}. 

\begin{figure}[htb!]
\begin{center}
\includegraphics[width=0.6\columnwidth]{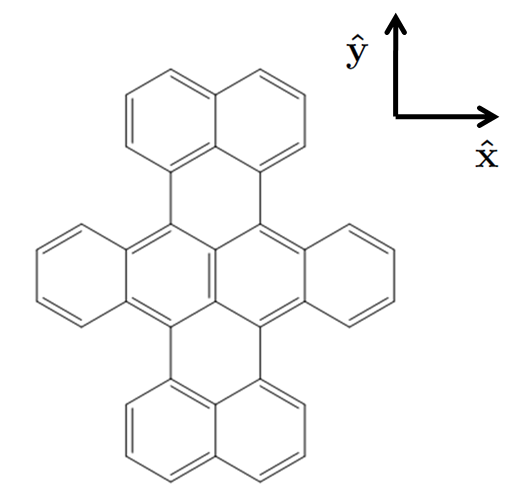}
\caption{A cartoon representation of DBT omitting  the hydrogen atoms. The two axes, labeled $\hat{\mbf{y}}$ and $\hat{\mbf{x}}$, are used to facilitate discussions. }
\label{dbtmolfig}
\end{center}
\end{figure}

Dibenzoterrylene (DBT) is an organic material that has garnered a lot of interest as a possible SPS \cite{C.Toninelli2010,Y.Tian2012,P.Siyushev2014, A.A.L.Nicolet2007,A-M.Boiron1996,C.Polisseni2016,J.Hwang2011,D.Wang2017,N.R.Verhart2016,F.Jelezko1996,C.Hofmann2005,A.A.L.Nicolet2013,S.Grandi2016}. A cartoon representation of DBT is shown in Fig. \ref{dbtmolfig}. Typically, DBT is deposited in an anthracene (Ac) matrix \cite{C.Polisseni2016, A.Makarewicz2012}, primarily to guard against oxidation and photobleaching, as these processes limit the photostability of the system. DBT has a purely electronic zero phonon line (ZPL) around 785 nm. At low temperatures, the phonon induced dephasing of the transition dipole of the ground state to the first bright excited state, which is labeled $S_{1}$, vanishes \cite{B.Kozankiewicz2014}; the spectral line width of this transition is then limited only by the radiative lifetime, and DBT can act as a two level system, similar to a trapped atom \cite{J.-B.Trebbia2009}.

%

The development of DBT as a SPS requires a detailed understanding of its electronic states. In particular, a disadvantage of using DBT as a SPS is its inter-system crossing (ISC), where the population of a singlet state is funneled to a triplet state \cite{C.Toninelli2010}. The rate at which ISC proceeds is exponentially suppressed by the energy difference between the singlet state and the triplet state \cite{R.Englman1970,Marian2012,Coyle1989}, and therefore the energy of the triplet state is important in considering the use of DBT as a SPS. There is no consensus on the energy of the first triplet state of DBT, primarily due to experimental limitations. Some researchers have calculated this triplet state to have an energy as low as 0.23 $\,e$V above the ground state \cite{I.Deperasinska2010}. This calculated energy is very different from that of the triplet states in the -acenes, materials of similar structure to DBT, which typically have triplet energies on the order of 1 $\,e$V above the ground state \cite{M.B.Smith2013,Z.S.Sadeq2015}. 

There are multiple strategies for calculating the energy of the electronic states of DBT, including Density Matrix Renormalization Group (DMRG)  \cite{C.Raghu2002a,J.Hachmann2007, I.Hagymasi2016} and Density Functional Theory (DFT) \cite{I.Deperasinska2010,J.Fabian2002,L.Serrano-Andres2005}. While there have been DMRG studies on the electronic states of systems with similar structure to DBT \cite{C.Raghu2002a, J.Hachmann2007}, no calculations have been performed on DBT; with techniques such as DFT, calculations can exhibit large variations in the predicted energies \cite{I.Deperasinska2010,Y.X.Yao2015}. Crucially, these approaches make it difficult to get a simple picture of the electronic behavior in the excited states. Earlier \cite{Z.S.Sadeq}, we used the Pariser-Parr-Pople (PPP) model \cite{Barford2005,R.Pariser1953,R.Pariser1953a,Pople1954} to describe the electronic and optical properties of graphene flakes of similar size and structure to DBT. The PPP model has been successfully implemented to study a range of carbon based materials, from pentacene \cite{C.Raghu2002a} to graphene flakes \cite{Z.S.Sadeq,J.A.Verges2015}. In this paper, we use this same model to elucidate the electronic excited states of DBT. The attractive feature of our approach is its ability to provide a simple physical picture of the electron behavior in these states. We calculate the linear and nonlinear optical absorption of DBT and the electron densities involved in several bright transitions. We demonstrate that there is no other competing one-photon absorption process near the $S_{1}$ absorption, and that there are no optical processes that might hinder the application of DBT as a SPS.


This paper is written in four parts: In Section \ref{sec:mod} we discuss the model used to describe the electronic states of DBT, in Section \ref{sec:abs-spec} we compute the one-photon and two-photon absorption of DBT and the spatial distribution of the electrons involved in several bright transitions, and in Section \ref{sec:conc} we present our conclusions. 

\section{\label{sec:mod} Model and Methods}
We model the $p_{z}$ electrons in DBT using the Pariser-Parr-Pople (PPP) Hamiltonian \cite{C.Raghu2002a,Z.S.Sadeq,Barford2005,R.Pariser1953,R.Pariser1953a,Pople1954},
\beq
H = H_{TB}+ H_{Hu} + H_{ext},
\label{eq-totham}
\eeq
where $H_{TB}$ is the tight-binding Hamiltonian, $H_{Hu}$ is the Hubbard Hamiltonian, and $H_{ext}$ is the extended Hubbard Hamiltonian,
\beq
&& H_{TB} = - \sum_{\la i,j \ra, \sigma} t_{ij} c^{\dg}_{i\sigma} c_{j\sigma} \label{TBeq}, \\
&& H_{Hu} = U \sum_{i} n_{i\ua} n_{i\da} \label{hubb}, \\
&& H_{ext} = \frac{1}{2} \sum_{\substack{i \neq j \\ \sigma \sigma'}} V_{ij} \left( n_{i\sigma} - \frac{1}{2} \right) \left(n_{j\sigma'} - \frac{1}{2} \right). \label{exthubb}
\eeq
Here $\sigma$ is a spin label, $i$ and $j$ are site labels, and the angular brackets indicate sums over nearest neighbors only.
The hopping parameter is set to $t_{ij} = 2.66 \,e$V for the $\pi$ conjugated bonds,
and $t_{ij} = 2.22 \,e$V for the single bonds  \cite{Kundu2009,J.Hachmann2007}. 
The fermion creation and annihilation operators are denoted respectively by $c^{\dg}_{i\sigma}$ and $c_{i\sigma}$, so the electron number operator for spin $\sigma$ and site $i$ is $n_{i\sigma} = c^{\dg}_{i\sigma} c_{i\sigma}$. 
The Hubbard repulsion parameter is set to $U = 5.88$ $\,e$V for all calculations; this choice of $U$ ensures that the first singlet transition energy matches the experimental value. The value of $U$ we use is similar to values that have  been used to model the $p_{z}$ electrons in other organic systems \cite{Barford2005,J.A.Verges2015,Z.S.Sadeq}. We approximate the long-range Coulomb repulsion by the Ohno interpolation \cite{Barford2005},
\beq
V_{ij} = \frac{U}{\sqrt{1 + \left( 4\pi \epsilon_{0} U  \epsilon r_{ij} / e^2 \right)^{2}}},
\eeq
where $U$ is the on-site repulsion parameter, $\epsilon$ is a screening parameter, $r_{ij}$ is the distance between sites $i$ and $j$, $e = -|e|$  is the electronic charge, and $\epsilon_{0}$ is the vacuum permittivity.  We set $\epsilon = 5$ for all calculations, as this value has been used to study similar systems such as graphene flakes \cite{J.A.Verges2015,Z.S.Sadeq}.

We first consider the Hartree-Fock (HF) approximation for the PPP Hamiltonian (\ref{eq-totham}),
\begin{widetext}
\begin{align}
H^{HF} = & - \sum_{\la i,j \ra, \sigma} t_{ij} c^{\dg}_{i\sigma} c_{j\sigma}  +  U \sum_{i}  \left( \la n_{i\ua} \ra n_{i\da} + \la n_{i\da} \ra n_{i\ua} - \la n_{i\ua} \ra \la n_{i\da} \ra - \la c^{\dg}_{i\ua} c_{i\da} \ra c^{\dg}_{i\da} c_{i\ua} - \la c^{\dg}_{i\da} c_{i\ua} \ra  c^{\dg}_{i\ua} c_{i\da} + \la  c^{\dg}_{i\ua} c_{i\da} \ra \la  c^{\dg}_{i\da} c_{i\ua} \ra \right) \nonumber 
\\
& + \sum_{i \neq j} V_{ij} \left( n_{i} \la n_{j} \ra  - n_{i} 
- \frac{1}{2} \la n_{i} \ra \la n_{j} \ra + \frac{1}{2}
- \frac{1}{2} \sum_{\sigma\sigma'} \la c^{\dg}_{i\sigma} c_{j\sigma'} \ra c^{\dg}_{j\sigma'} c_{i\sigma} + \la c^{\dg}_{j\sigma'} c_{i\sigma} \ra c^{\dg}_{i\sigma} c_{j\sigma'} - \la c^{\dg}_{i\sigma} c_{j\sigma'} \ra \la c^{\dg}_{j\sigma'} c_{i\sigma}  \ra \right), \label{DBTHFeqs}
\end{align}
\end{widetext}
The derivation of these equations is well known  \cite{J.A.Verges2015,H.Bruus2004,G.F.Giuliani2005,Pines1977}. We diagonalize $H^{HF}$ and solve for the expectation values self consistently, using the tight-binding (\ref{TBeq}) eigenstates as an initial guess. The HF Hamiltonian (\ref{DBTHFeqs}) can then be written in diagonal form as 
\beq
H^{HF} = \sum_{m \sigma} \hbar\omega_{m\sigma} C^{\dg}_{m\sigma} C_{m\sigma},
\eeq
where $\hbar \omega_{m\sigma}$ are the eigenvalues associated with the single particle states, and $C^{\dg}_{m\sigma}$ is the corresponding creation operator for a particle with spin $\sigma$.
The operators $C^{\dg}_{m\sigma}$ and $C_{m\sigma}$ can be written in terms of the site basis operators as
\beq
&& C^{\dg}_{m\sigma} = \sum_{i} M_{m\sigma,i} c^{\dg}_{i\sigma}, \\
&& C_{m\sigma} = \sum_{i} M^{*}_{m\sigma,i} c_{i\sigma},
\eeq
where $M_{m\sigma,i}$ is the amplitude associated with the state $m$ at site $i$, and is typically non-zero for all $i$. Since $M_{\sigma}$ is a unitary matrix, the HF quasiparticle operators obey the fermionic anticommutation relations $\{ C_{m\sigma}, C^{\dg}_{m'\sigma'} \} = \delta_{mm'}\delta_{\sigma\sigma'}$. The single particle states obtained from solving the HF equations with paramagnetic expectation values are then used to construct the HF ground state. Single-particle states that are filled in the HF ground state are denoted as ``valence", and the  unfilled ones are denoted as ``conduction". We then rewrite the total Hamiltonian (\ref{eq-totham}) in an electron-hole basis. In the electron-hole basis, the HF electron creation operator is designated by $a^{\dg}_{m\sigma}$, and the HF hole creation operator is designated by  $b^{\dg}_{m'\sigma}$, so 
\beq
a_{m\sigma} = C_{m\sigma}, 
 \quad  \quad
b^{\dg}_{m'\sigma} = C_{m'\tilde{\sigma}}, \label{hfhole}
\eeq
where $\tilde{\sigma}$ is the opposite spin of $\sigma$, $m$ indicates a conductance state, and $m'$ indicates a valence state. 

To solve for the many body wavefunctions of the system, we restrict the many body Hamiltonian (\ref{eq-totham}) to a set of states following the configuration interaction (CI) method: We employ a basis consisting of the HF ground state, and HF single and double excitations. Upon diagonalization of the many body Hamiltonian, the CI ground state and the CI excited states are superpositions of the HF ground state and the HF excited states. We then diagonalize the many body Hamiltonian (\ref{eq-totham}) restricted to the selected states to obtain the many body wavefunctions. The details of the electron hole basis, and the CI strategy used to solve for the many body wavefunctions, can be found in our earlier work \cite{Z.S.Sadeq}. For the rest of this paper, we shall refer to the HF single particle states as ``modes'', and we shall refer to states that result from the CI calculation simply as ``states''.


\subsection*{States and Transitions of Interest}
We label the first four bright excited states in ascending energy as the $S_{n}$ states, where $n = \left\{ 1,2,3,4\right\} $. For the application of DBT as a SPS, the transition from the ground state to the $S_{1}$ state, denoted $GS \rightarrow S_{1}$, is the transition of interest; the relaxation of the excitation from the $S_{1}$ state to the $GS$ is the source of single photons \cite{C.Polisseni2016}. We label the lowest energy two-photon active state as the $2LH$ state; this state is primarily composed of a HF double excitation that excites two electrons from the highest occupied HF mode to the lowest unoccupied HF mode. The next two two-photon active states in order of increasing energy are labeled as $S_{D_{1}}$ and $S_{D_{2}}$; these states are composed mainly of HF single excitations. We denote the first triplet state as $T_{1}$. The electronic population in the $S_{1}$ state can decay to the $T_{1}$ state via ISC \cite{Coyle1989}; the energy of the $T_{1}$ state is important as it represents a source of loss for the SPS application. The energies of these states above the ground state are shown in the plot in Fig. \ref{DBT_abs_spec} (b).  

\subsection*{Optical Response}
The number operator for a particular site $i$ is
\beq
n_{i} = \sum_{\sigma} c^{\dg}_{i\sigma} c_{i\sigma},
\label{eq-dop}
\eeq
and in the electron-hole basis \cite{Z.S.Sadeq}, it is written as  
\begin{align}
 n_{i} = & \sum_{mm'\sigma} \Gamma_{mm'\sigma,i} \left(a^{\dg}_{m\sigma} a_{m'\sigma} - b^{\dg}_{m'\sigma} b_{m\sigma}  \right) \nonumber \\
& + \sum_{mm'\sigma} \Gamma_{mm'\sigma,i} \left( a^{\dg}_{m\sigma} b^{\dg}_{m'\tilde{\sigma}} + b_{m\tilde{\sigma}}a_{m'\sigma} \right) + \sum_{m\sigma} \Gamma_{mm\sigma,i},
\end{align}
where we have defined
\beq
\Gamma_{mm'\sigma,i} = M_{m\sigma,i} M^{*}_{m'\sigma,i}.
\eeq
The dipole moment operator of the system is approximated as
\beq
\bso{\mu}  = \sum_{i} e\mbf{r}_{i} \left( n_{i}-1 \right),
\eeq
where $\mbf{r}_{i}$ is the location of site $i$. Transforming into the electron-hole basis \cite{Z.S.Sadeq}, we have
\begin{align}
\bso{\mu}  = & \sum_{m\sigma} \bso{\mu}_{mm\sigma}  - e\sum_{i} \mbf{r}_{i} +  \sum_{mm'\sigma} \bso{\mu}_{mm'\sigma} a^{\dg}_{m\sigma} a_{m'\sigma}  \nonumber \\
& - \sum_{mm'\sigma} \bso{\mu}_{mm'\sigma} b^{\dg}_{m'\sigma} b_{m\sigma} +  \sum_{mm'\sigma} \bso{\mu}_{mm'\sigma} a^{\dg}_{m\sigma} b^{\dg}_{m'\tilde{\sigma}} \nonumber \\
& + \sum_{mm'\sigma} \bso{\mu}_{mm'\sigma} b_{m\tilde{\sigma}} a_{m'\sigma},
\end{align}
where  
\beq
\bso{\mu}_{mm'\sigma} = \sum_{i} e \mbf{r}_{i} \Gamma_{mm'\sigma,i}.
\eeq
The matrix elements of the dipole moment operator arise in calculating the optical absorption of the material, to which we now turn.

\subsubsection*{Linear Response}

The one-photon absorption spectrum can be determined from the imaginary component of the first order polarizability of the system \cite{Boyd2008}; assuming the system is initially in the ground state, the imaginary component of the first order polarizability \cite{Boyd2008a} is given by
\beq
\text{Im} \left( \alpha^{(1)}_{kl}(\omega) \right) = \frac{1}{\epsilon_{0} \hbar} \sum_{m} \frac{ \gamma_{mg} \mu_{gm}^{k}\mu_{mg}^{l}}{ \left(\omega_{mg}-\omega\right)^{2}+\gamma_{mg}^{2}},
\label{eq-imchi1}
\eeq
where $\omega$ is the frequency, $k,l$ are Cartesian components, $\epsilon_{0}$ is the vacuum permittivity, $\bso{\mu}_{mg}$ is the matrix element of the dipole moment operator between the ground state and the state $m$, $\omega_{mg}$ is the frequency difference between the state $m$ and the ground state, and $\gamma_{mg}$ is the frequency broadening associated with the transition from the ground state to the state $m$. The one-photon absorption coefficient measured in experiments is proportional to the imaginary component of the linear susceptibility of the system \cite{Wei2015}, which can be obtained from the linear polarizability \cite{Boyd2008}. We report the predicted strength of each one-photon absorption peak in terms of the oscillator strength associated with the corresponding transition. The predicted oscillator strength of the absorption peak due to the transition from the ground state to the state $Y$  \cite{Boyd2008d}, denoted by $GS \rightarrow Y$, is
\beq
f_{Yg} = \frac{2m_{e}\omega_{Yg} \left| \bso{\mu}_{Yg} \right|^{2}}{3\hbar e^{2}},
\eeq
where $m_{e}$ is the electron mass. 

We define the ``spatial profile'' of the transition $GS \rightarrow Y$ as a function $\mathcal{T}_{Y;i}$ of site $i$ given by
\beq
\mathcal{T}_{Y;i} = \la Y | n_{i} | GS \ra,
\eeq
where $GS$ is the ground state, $i$ is the site, and $Y$ is an excited state. This quantity is related to the matrix element of the dipole moment operator  between the ground state and the state $Y$, the ``transition dipole moment'',
\beq
\la Y | \bso{\mu} | GS \ra = \sum_{i} e\mbf{r}_{i} \mathcal{T}_{Y;i},
\eeq
where the sum is over all sites $i$.

%

\subsubsection*{Nonlinear Response}

The two-photon absorption spectrum can be determined from the imaginary component of the third order polarizability of the system \cite{Boyd2008}; assuming the system is initially in the ground state, the largest contribution to the third order polarizability for two-photon absorption \cite{Boyd2008b} is given by
\begin{widetext}
\beq
\alpha^{(3)}_{klop} \left(\omega; \omega, \omega,-\omega \right) \approx \frac{1}{\epsilon_{0} \hbar^{3}}\mathcal{P}_{I} \sum_{vnm} \frac{\mu^{k}_{gv} \mu^{l}_{vn} \mu^{o}_{nm} \mu^{p}_{mg} }{\left(\omega_{vg} - \omega - i\gamma_{vg} \right) \left(\omega_{ng} - 2\omega - i\gamma_{ng} \right) \left(\omega_{mg} - \omega - i\gamma_{mg} \right)} ,
\label{chi3_form}
\eeq
\end{widetext}
where $\mathcal{P}_{I}$ is the permutation operator in the set of distinct frequencies $\left\{ \omega,\omega,-\omega\right\} $, $\hbar\omega_{vm}$ is the energy difference between states $v$ and $m$, and $\gamma_{vg}$ is the frequency broadening associated with the transition $GS \rightarrow v$. The two-photon absorption coefficient measured in experiments is proportional to the imaginary component of the third order susceptibility of the system \cite{Wei2015}, which can be obtained from the third order polarizability \cite{Boyd2008}. The predicted strength of the two-photon transition from the ground state to the state $Z$ is given by
\beq
B_{klop} \left( GS \rightarrow Z \right) =  \frac{\pi}{2\epsilon_{0} \hbar^{3}}\sum_{vm} \frac{ \mu^{k}_{gv} \mu^{l}_{vZ} \mu^{o}_{Zm} \mu^{p}_{mg}}{\left( \omega_{vg} - \overline{\omega} \right) \left( \omega_{mg} - \overline{\omega} \right)},
\label{eq:Bklop}
\eeq
where $\overline{\omega} = \omega_{Zg} / 2$. A derivation of $B_{klop} \left( GS \rightarrow Z \right)$ is presented in Appendix \ref{dbt-app}.

\begin{figure*}[htb!]
\begin{tabular}{c}
\includegraphics[width=1\textwidth]{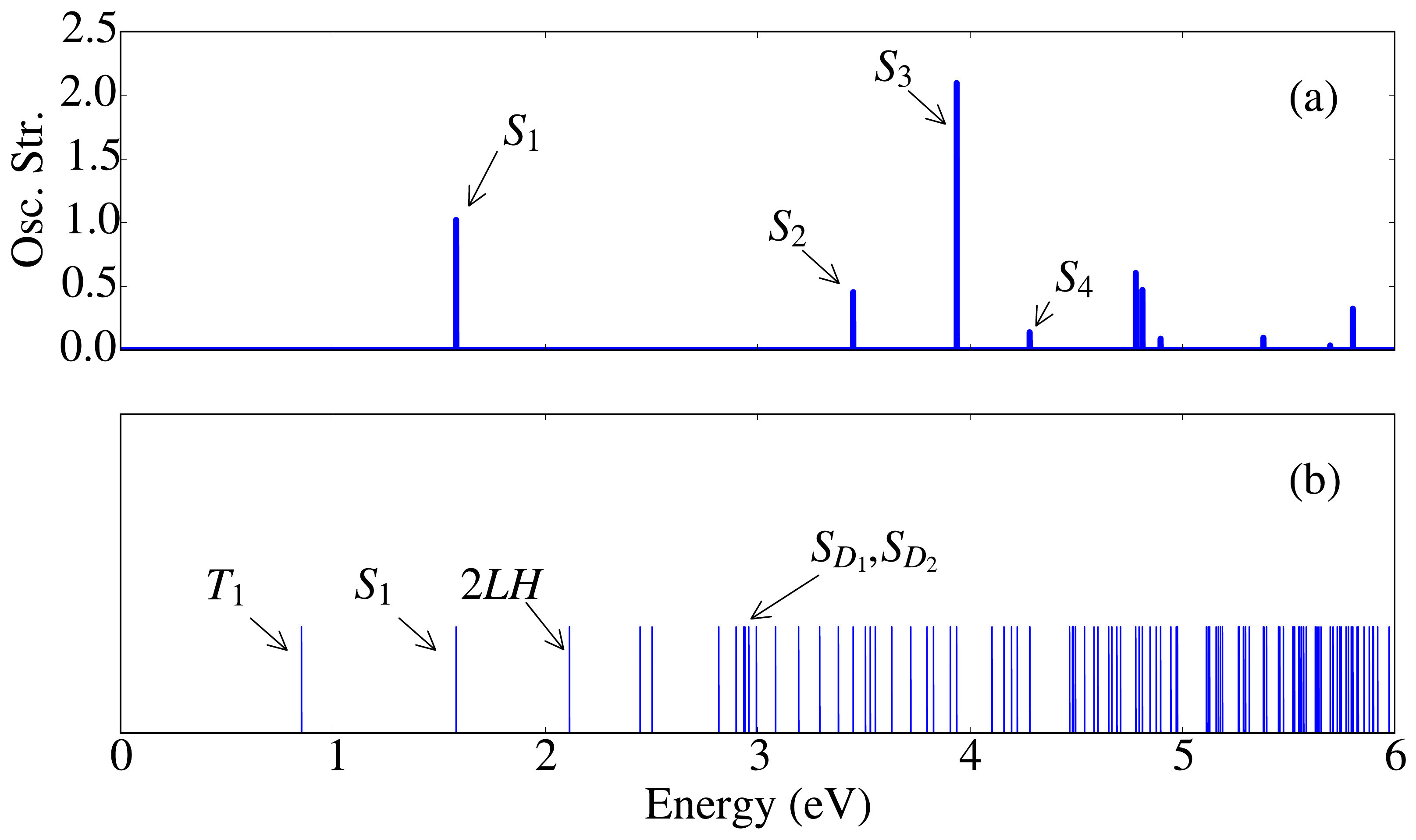}
\end{tabular}
\caption{ (a) A plot of the oscillator strengths of the bright transitions, and (b) the energies of the excited states above the ground state of DBT. The parameters of our calculation are set such that the $S_{1}$ state is 1.58 $\,e$V above the ground state in energy. Our calculation predicts that the $T_{1}$ state is 0.85 $\,e$V above the ground state in energy. The plot of the energies of the excited states above the ground state indicates that there are no other states below the $S_{1}$ state in energy besides the $T_{1}$ state. The first absorption peak, due to $GS \rightarrow S_{1}$, has an associated transition dipole moment which is polarized along the $\hat{\mbf{y}}$ axis. There are several other high energy absorption peaks, the strongest of which is due to $GS \rightarrow S_{3}$; this transition has a transition dipole moment which is polarized along the $\hat{\mbf{x}}$ axis. However, these bright states are all more than $3$ $\,e$V above the ground state in energy, which is around twice the energy of the $GS \rightarrow S_{1}$ transition. The axes are shown in Fig. \ref{dbtmolfig}. }
\label{DBT_abs_spec}
\end{figure*}

\begin{figure}[htb!]
\begin{tabular}{c}
\includegraphics[width=1.0\columnwidth]{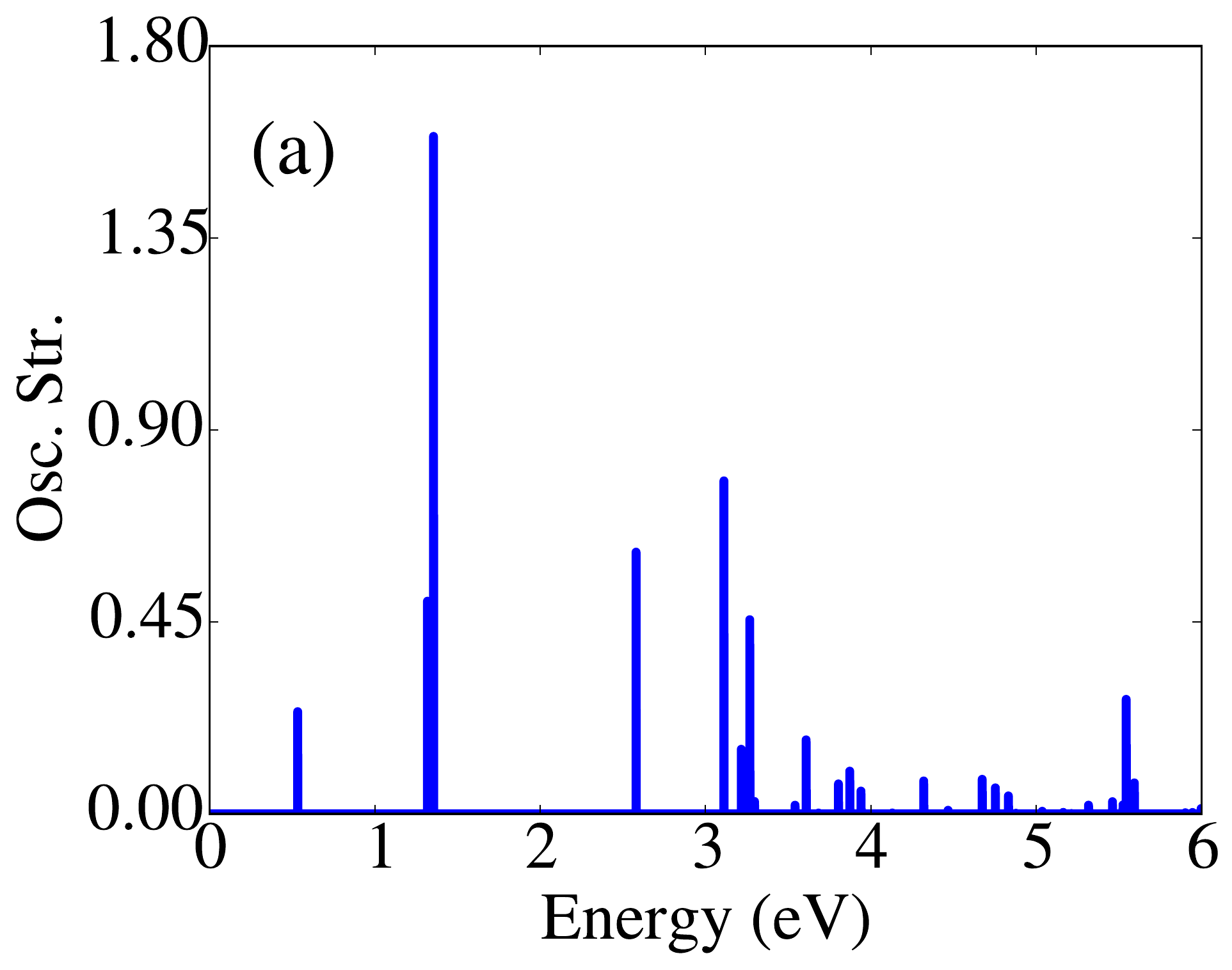} \\
\includegraphics[width=1.0\columnwidth]{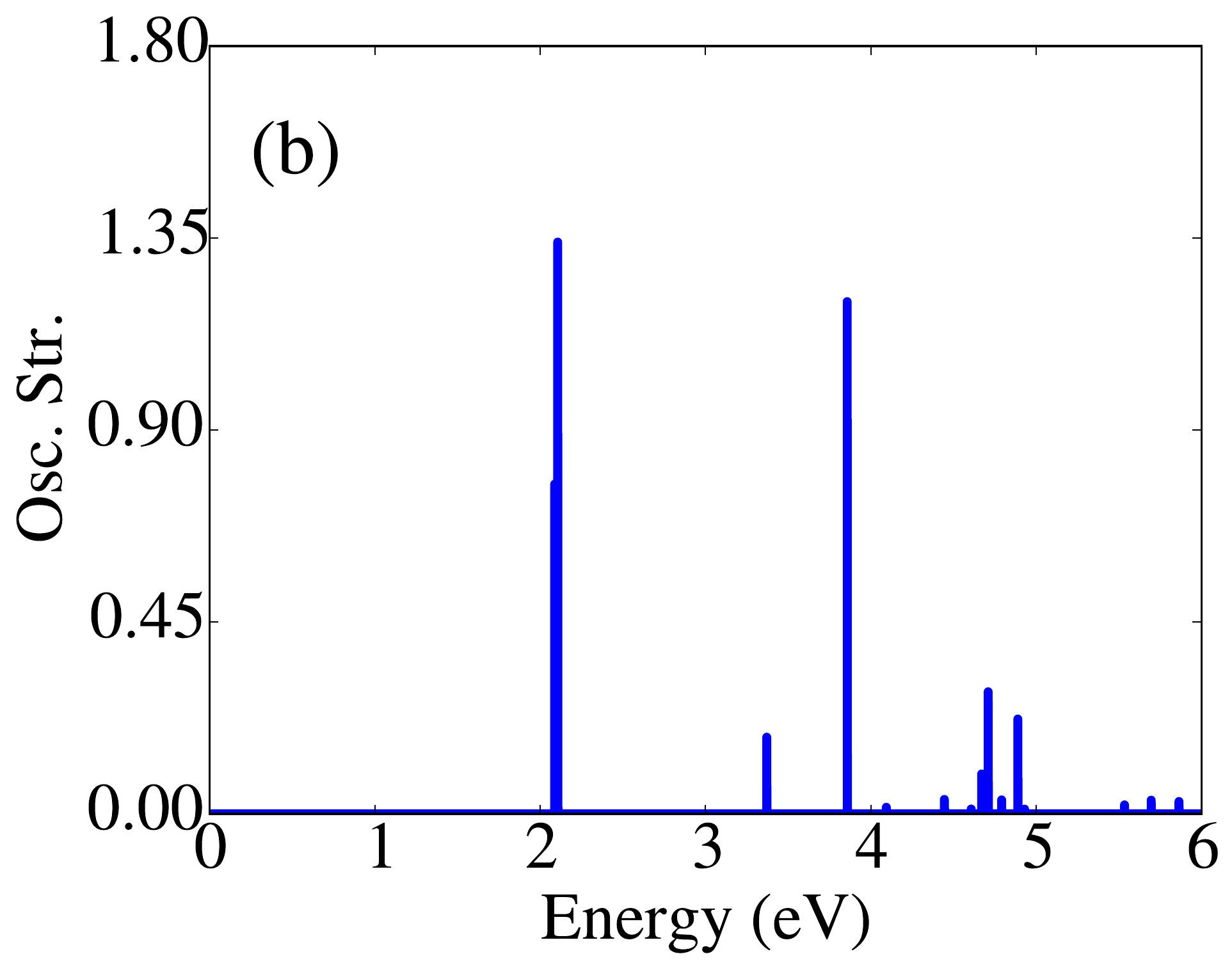}
\end{tabular}
\caption{The oscillator strengths of the bright transitions with (a) the first singlet state $S_{1}$, and (b) the first triplet state $T_{1}$, as the initial state. Further absorption from the $S_{1}$ or $T_{1}$ state does not fall within the energy range of the transition from the ground state to the $S_{1}$ state.}
\label{DBT_abs_specS1}
\end{figure}

\begin{figure*}[htb!]
\begin{tabular}{rrrr}
\includegraphics[width = 0.25\textwidth]{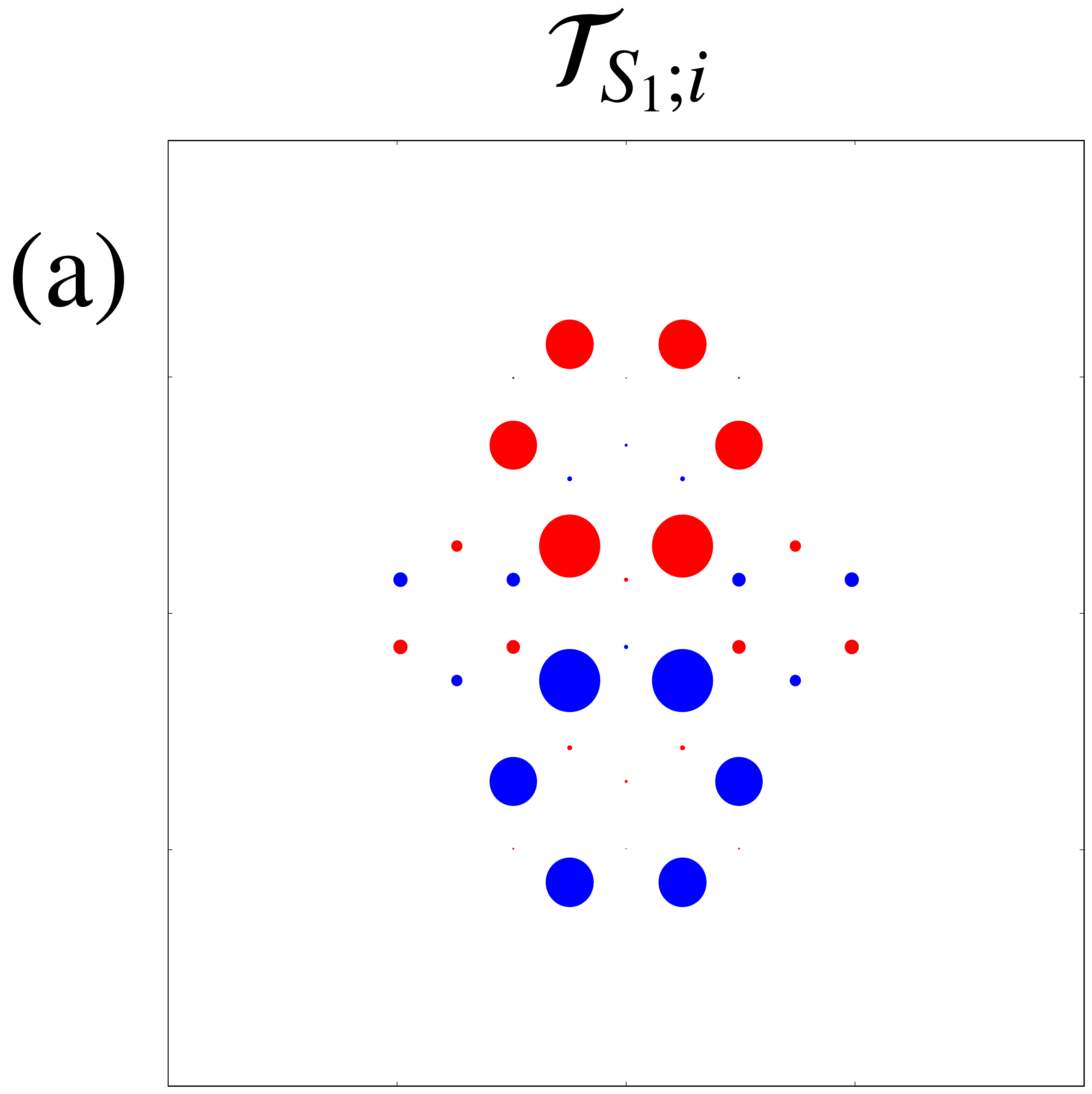} &
\includegraphics[width = 0.25\textwidth]{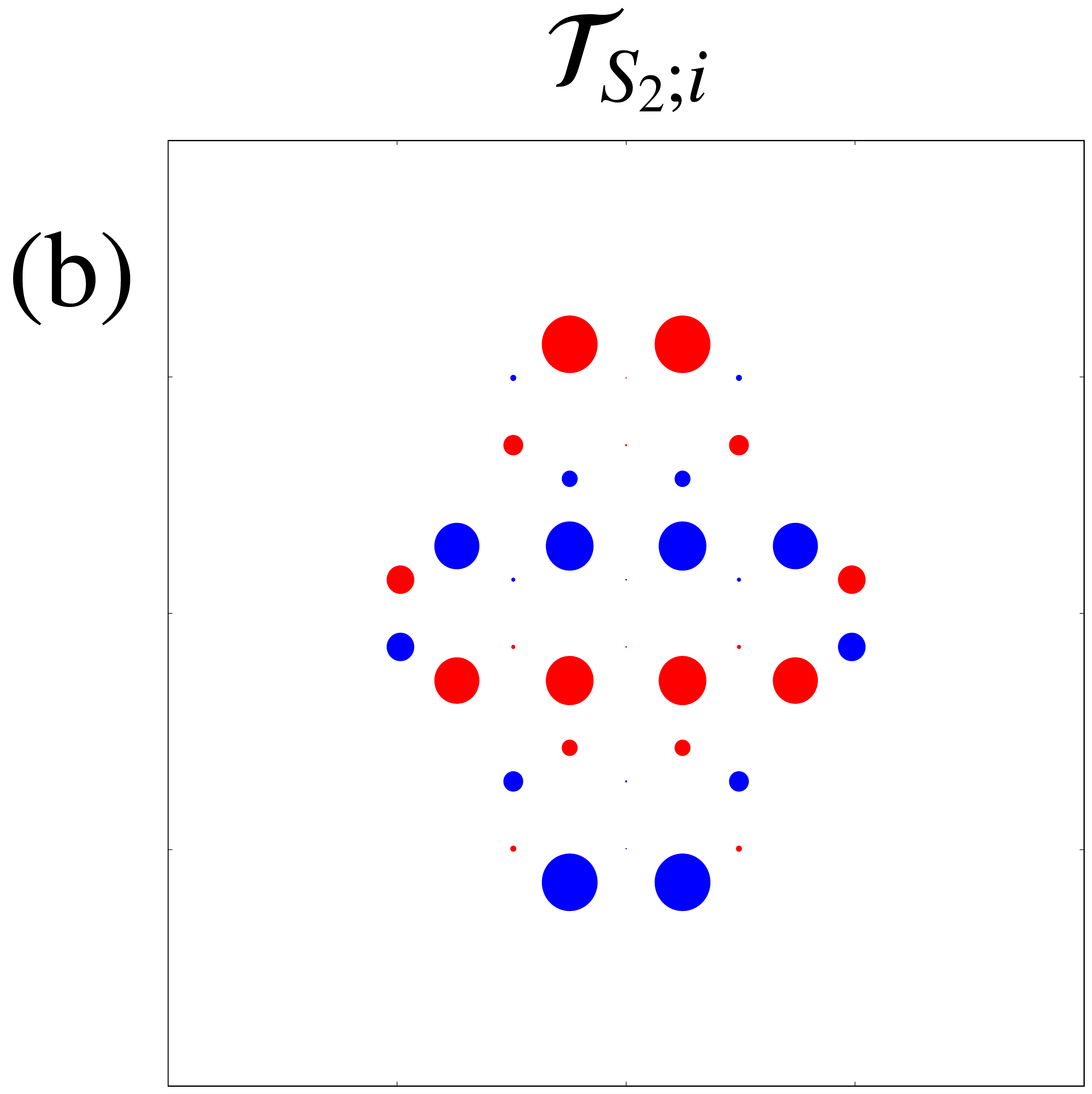} &
\includegraphics[width = 0.25\textwidth]{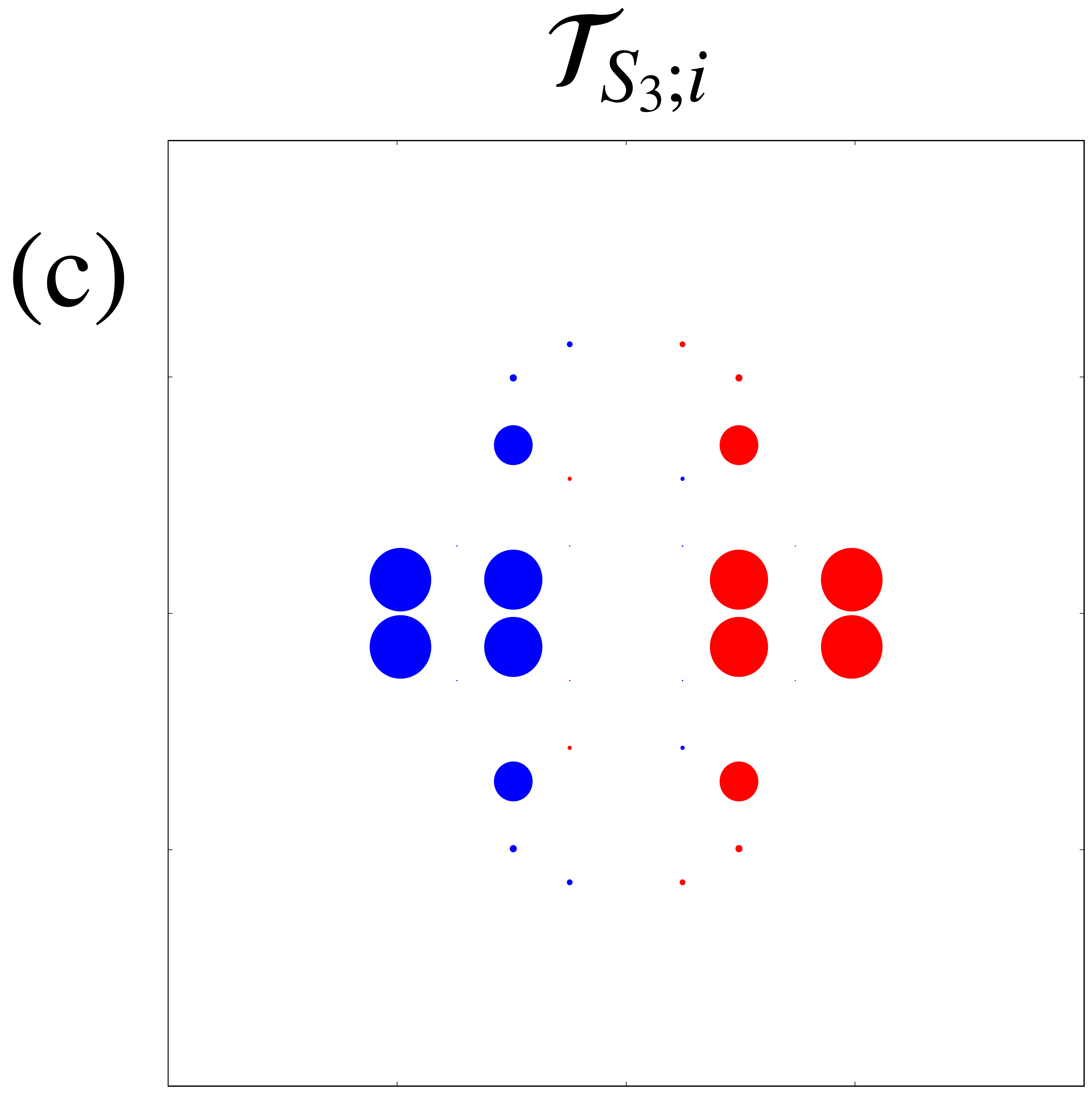} & 
\includegraphics[width = 0.25\textwidth]{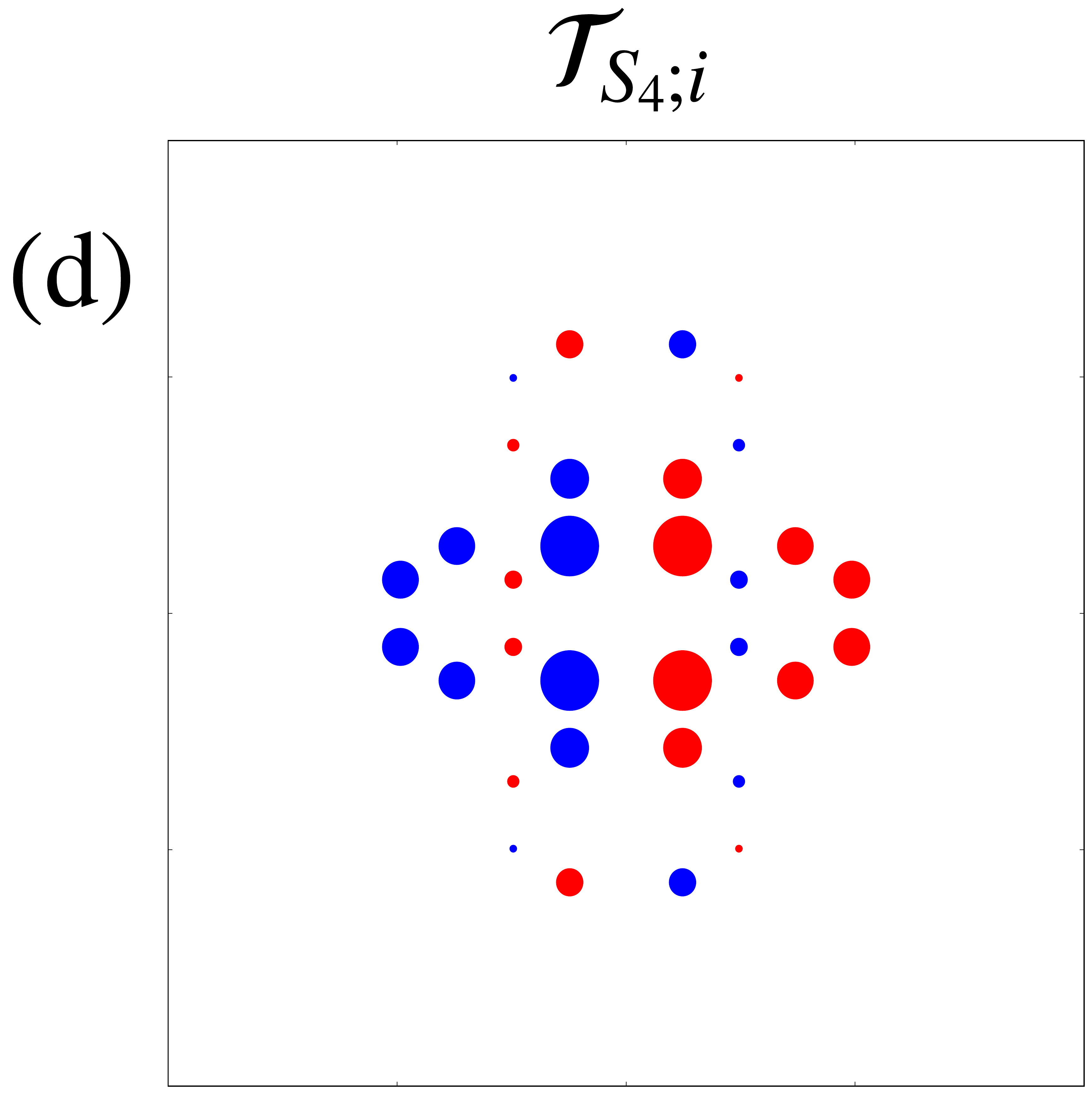} 
\end{tabular}
\caption{Plots of (a) $\mathcal{T}_{S_{1};i}$, (b)  $\mathcal{T}_{S_{2};i}$, (c)  $\mathcal{T}_{S_{3};i}$, and (d)  $\mathcal{T}_{S_{4};i}$. We place a circle at the location of each site $i$; the area of each circle indicates the magnitude of $\mathcal{T}_{Y;i}$, and the color indicates whether it is positive (red) or negative (blue). The convention used for labeling these states follows energetic order, i.e. $S_{1}$ is the lowest energy state, $S_{2}$ is the second lowest energy state and so on.  The transitions that are polarized along the $\hat{\mbf{y}}$ axis have electron concentration extended throughout the system, while those polarized along the $\hat{\mbf{x}}$ axis have electron concentration primarily in the middle of the system; the axes are shown in Fig. \ref{dbtmolfig}.}
\label{DBT_td}
\end{figure*}

\section{\label{sec:abs-spec} Optical Absorption of DBT}

\subsection{Linear Absorption Spectrum and Spatial Profiles of Bright Transitions}


In Fig. \ref{DBT_abs_spec} (a), we plot the oscillator strengths of the bright transitions of DBT assuming the system is initially in the ground state. Recall that the strength of the Coulomb repulsion parameter $U$ was set so the energy of $GS \rightarrow S_{1}$ is 1.58 $\,e$V, in agreement with the experimental value. The radiative line width of $GS \rightarrow S_{1}$ in vacuum \cite{Boyd2008c} is given by 
\beq
\Gamma^{\text{vac}}_{S_{1}} = \frac{\omega^{3}_{S_{1}g} \vert \bso{\mu}_{S_{1}g} \vert^{2}}{3\pi\epsilon_{0}\hbar c^{3}},
\eeq
where $c$ is the speed of light, $\omega_{S_{1}g}$ is the frequency difference between the $S_{1}$ state and the ground state, and $|\bso{\mu}_{S_{1}g}|$ is the magnitude of the matrix element of the dipole moment operator between the ground state and the $S_{1}$ state; we find $|\bso{\mu}_{S_{1}g}| = 13.1 \text{ Debye}$. Due to local field effects, the radiative line width of $GS \rightarrow S_{1}$ is modified when DBT is deposited in an anthracene matrix. There is some controversy as to which is the most appropriate model to describe the local field effects when calculating the radiative line width of emitters embedded in a homogeneous medium  \cite{K.Dolgaleva2007,K.Dolgaleva2012}.  The real cavity model describes local field effects when the emitters, in this case the DBT molecules, enter the medium as dopants \cite{K.Dolgaleva2007,K.Dolgaleva2012}. Accounting for local field effects using the real cavity model, the radiative line width of the $GS \rightarrow S_{1}$ transition is
\beq
\Gamma^{\text{RC}}_{S_{1}} = n_{\text{eff}} \left( \frac{3n^{2}_{\text{eff}}}{2n^{2}_{\text{eff}} + 1} \right)^{2} \Gamma^{\text{vac}}_{S_{1}},
\label{eq-dbt-ac-linewidth}
\eeq
where $n_{\text{eff}}$ is the effective refractive index of the material. Since the concentration of DBT in the anthracene matrix is extremely dilute \cite{C.Polisseni2016,K.D.Major2015}, we take $n_{\text{eff}}$ to be the refractive index of anthracene. Using Eq.~(\ref{eq-dbt-ac-linewidth}), we calculate the radiative line width of $GS \rightarrow S_{1}$ to be 40 MHz; a complete neglect of local field corrections leads to a predicted radiative line width of 30 MHz.  Measurements of the homogeneously broadened radiative line width of the $GS \rightarrow S_{1}$ transition range from 30-40 MHz \cite{A.A.L.Nicolet2007,J.-B.Trebbia2009,S.Grandi2016}. These measurements were carried out at cryogenic temperatures, at which the radiative line width is generally assumed to be limited only by the excited state lifetime \cite{J.-B.Trebbia2009,T.Basche2007}. 
Given that our calculated radiative line width is in the range of  reported experimental values, and assuming the experimental value of the radiative line width is indeed limited only by the excited state lifetime, then the magnitude of our calculated $GS \rightarrow S_{1}$ transition dipole moment is also consistent with the experimental values.

In our calculations, the lowest triplet state, $T_{1}$, has energy 0.85 $\,e$V above the ground state, which is about half the energy of the first singlet excited state above the ground state. The energy of the triplet state is calculated to be greater than the corresponding value calculated by Deperasinska \textit{et al.} \cite{I.Deperasinska2010} by 0.62 $\,e$V. Their calculation predicts that the $T_{1}$ state is 0.23 $\,e$V above the ground state; however, they point out that the method they use to calculate the energies of the excited states exhibits an average deviation of $0.4$ $\,e$V between the calculated energies and the experimental values for small molecules, and gives less accurate results for systems as large as DBT \cite{I.Deperasinska2010,J.Fabian2002,L.Serrano-Andres2005}. The energies of the lowest triplet excited states in the -acene series, which are similar in structure to DBT, are approximately 1 $\,e$V above the ground state \cite{Z.S.Sadeq2015}; in the -acene series, the energy above the ground state of the lowest triplet excited state is approximately half of the energy above the ground state of the lowest singlet excited state, as we find for DBT. The energy of the triplet state in DBT has yet to be experimentally determined. Absorption to the other excited states requires photon energies greater than 3 $\,e$V. Our calculations indicate that there are no other absorption peaks close in energy to the absorption peak due to $GS \rightarrow S_{1}$, and we predict that there are no competing linear optical processes that might reduce the efficacy of the application of this material as a SPS. The plot of the energies of the excited states above the ground state, shown in Fig. \ref{DBT_abs_spec} (b), indicates that there are no other excited states that are lower in energy than the $S_{1}$ state, except for the $T_{1}$ state. Thus  from our calculations the only other major relaxation channel is the ISC to the $T_{1}$ state.



To consider the possible significance of sequential absorption from the ground state, we investigate the oscillator strengths of the bright transitions with (a) $S_{1}$ and (b) $T_{1}$ as the initial state. We plot these oscillator strengths in Fig. \ref{DBT_abs_specS1}. It is clear that further absorption from the $S_{1}$ state either occurs at photon energies below (less than $0.5$ $\,e$V) or above (greater than $2$ $\,e$V) the energy of $GS \rightarrow S_{1}$. If the excitation decays to the triplet state, due to ISC for example, any further absorption from the triplet state occurs at photon energies that are much higher (greater than $2$ $\,e$V) than the energy of $GS \rightarrow S_{1}$. Our calculations indicate that there is no further absorption from the $S_{1}$ or $T_{1}$ state for the energy range of interest of the application of DBT as a SPS. 
%


Finally, we turn to the spatial profiles of the bright transitions from the ground state of DBT. We calculate $\mathcal{T}_{Y;i}$ for the first four bright excited states, and plot them in Fig. \ref{DBT_td}.  
%
The molecular axes are shown in Fig.~\ref{dbtmolfig}. The transition dipole moments associated with the first two optical transitions, $GS \rightarrow S_{1}$ and $GS \rightarrow S_{2}$, are polarized along the $\hat{\mbf{y}}$ axis. The spatial profile of $GS \rightarrow S_{1}$ has electron density extended across the entire system and leads to a very large transition dipole moment. The spatial profile of $GS \rightarrow S_{2}$ has electron concentration extended across the entire system, much like $\mathcal{T}_{S_{1};i}$, but has a weaker transition dipole moment. The transition dipole moments associated with the next two optical transitions, $GS \rightarrow S_{3}$ and $GS \rightarrow S_{4}$,  are polarized along the $\hat{\mbf{x}}$ axis. The spatial profile of $GS \rightarrow S_{3}$ has electron density primarily on the four rings in the middle of the system, and it has very little concentration on the top and bottom rings; it leads to a very large transition dipole moment. The spatial profile of $GS \rightarrow S_{4}$ has significant electron concentration in the middle of the system, and it has negligible electron concentration in the top and bottom rings of the system; it leads to a weaker transition dipole moment. For transitions with dipole moment polarized along the $\hat{\mbf{y}}$ axis, the electron density is extended across the entire system, but for transitions with dipole moment polarized along the $\hat{\mbf{x}}$ axis, the electron density is concentrated at the center of the system.

\subsection{Two-Photon Absorption Spectrum}

\begin{table}[htb!]
\begin{center}
\begin{tabular}{cccc}
    \hline
    Transition & $\hbar\omega$ ($\,e$V) & Strength ($\mu$m$^{5}$/V$^{2}$s) & Component  \\
    \hline
    $GS \rightarrow 2LH$ & 1.06  & 0.599  & $yyyy$  \\
    $GS \rightarrow S_{D_{1}}$  & 1.45 & 8.08 & $yxxy$  \\
    $GS \rightarrow S_{D_{2}}$  & 1.47 & 33.8 & $yyyy$ \\
    \hline
   \end{tabular}
   \caption{The lowest three two-photon transitions of DBT, their associated fundamental photon energies $\hbar\omega$, the associated integrated third order polarizability strengths, and the component of the third order polarizability tensor that exhibit the peaks.}
   \label{TPA_DBT}
\end{center}
    \end{table}

For the two-photon transitions of interest, we compute their associated fundamental photon energies, two-photon transition strengths, and the components of the third order polarizability tensor that exhibit the peaks. In Table \ref{TPA_DBT} we show the values of these quantities. The lowest energy TPA is due to $GS \rightarrow 2LH$, and arises from $\text{Im } (\alpha^{(3)}_{yyyy})$; the $2LH$ state is composed mainly of HF double excitations. The next two TPA peaks are due to $GS \rightarrow S_{D_{1}}$ (from $\text{Im } (\alpha^{(3)}_{yxxy})$) and $GS \rightarrow S_{D_{2}}$ (from $\text{Im } (\alpha^{(3)}_{yyyy})$); the $S_{D_{1}}$ state and the $S_{D_{2}}$ state are composed mainly of HF single excitations. The two-photon transitions $GS \rightarrow S_{D_{1}}$ and $GS \rightarrow S_{D_{2}}$ occur when the fundamental photon energy $\hbar\omega$ is close to the energy of the single photon transition $GS \rightarrow S_{1}$. The calculated strength of the TPA in DBT is in line with the TPA calculated in other conjugated organic systems \cite{Z.S.Sadeq2015,K.Aryanpour2014}. 




From the values of the calculated energies and the calculated absorption strengths, we argue that for the SPS application of DBT, the TPA should not compete with $GS \rightarrow S_{1}$ in any meaningful way. 

First we consider that a continuous wave (CW) laser is used to pump $GS \rightarrow S_{1}$ \cite{C.Polisseni2016}. Since the spectral width of CW lasers is usually less than $0.01 \, {\rm m}e$V, the spectrum of a CW laser centered at $\omega_{S_{1}g}$ will not contain the frequency components required to excite either $S_{D_{1}}$ or $S_{D_{2}}$. 



Second we consider the excitation by optical pulses, as done in a number of experiments \cite{J.-B.Trebbia2009,C.Toninelli2010, A.S.Clark2016}. For example, Toninelli \textit{et al.} \cite{C.Toninelli2010} used a Ti:Sapphire laser with a spectrum centered near $\omega_{S_{1}g}$ and a pulse duration of $120$ fs to excite $GS \rightarrow S_{1}$. The spectrum of these pulses has the necessary frequency components to excite the two-photon active transitions near $GS \rightarrow S_{1}$. To investigate the possible consequences of TPA, we use a perturbative treatment  \cite{Tannor2006} to calculate the one-photon absorption to the $S_{1}$ state and the two-photon absorption to the $S_{D_{2}}$ state. This approach is a generalization of  Eqs.~(\ref{eq-imchi1}) and (\ref{chi3_form}) respectively for pulsed pumping. We model the laser pulse as an unchirped Gaussian centered at $\omega_{S_{1}g}$ with an intensity full width at half maximum (FWHM) of $\tau$. Upon excitation of DBT, the population $\rho_{S_{1}}$ of the $S_{1}$ state at times after the pulse is
\beq
\rho_{S_{1}}=\left(\frac{\pi\vert\boldsymbol{\mu}_{S_{1}g}\vert^{2}}{\left(4\ln2\right)n_{\text{eff}}\epsilon_{0}c\hbar^{2}}\right)\tau^{2}I_{0},
\label{eq-s1-state-pulse}
\eeq
where $I_{0}$ is the peak intensity; $I_{0}$ can be written in terms of the pulse energy as
\beq
I_{0}=\frac{2\sqrt{\ln2}}{\sqrt{\pi}\tau}\frac{Q_{\text{pulse}}}{A_{\text{pulse}}},
\eeq
where $Q_{\text{pulse}}$ is the pulse energy and $A_{\text{pulse}}$ is the area of the laser spot.
The population $\rho_{S_{D_{2}}}$ of the $S_{D_{2}}$ state at such times is given by
\beq
\frac{\rho_{S_{D_{2}}}}{\rho_{S_{1}}}=\left(\frac{\vert\boldsymbol{\mu}_{S_{D_{2}}S_{1}}\vert^{2}\vert\mathcal{F}\vert^{2}}{\left(16\ln2\right)\pi n_{\text{eff}}\epsilon_{0}c\hbar^{2}}\right)\tau^{2}I_{0}.
\label{eq-ratio-pops}
\eeq
Here $|\bso{\mu}_{S_{D_{2}}S_{1}}|$ is the magnitude of the matrix element of the dipole moment operator between the $S_{1}$ state and the $S_{D_{2}}$ state; from our calculations $|\bso{\mu}_{S_{D_{2}}S_{1}}| = 17.6 \text{ Debye}$. The term $\mathcal{F}$ is
\begin{widetext}
\beq
\mathcal{F} = \intw \frac{\exp\left(-\frac{\tau^{2}}{4\ln2}\left(\omega-\omega_{S_{1}g}\right)^{2}\right)\exp\left(-\frac{\tau^{2}}{4\ln2}\left(\omega_{S_{D_{2}}g}-\omega-\omega_{S_{1}g}\right)^{2}\right)}{\left(\omega_{S_{1}g}-\omega-i(\Gamma_{S_{1}}/2)\right)} d\omega,
\eeq
\end{widetext}
where $\omega_{S_{D_{2}}g}$ is the frequency difference between the $S_{D_{2}}$ state and the ground state, and $\Gamma_{S_{1}}$ is the linewidth of $GS \rightarrow S_{1}$ given by (\ref{eq-dbt-ac-linewidth}). For $\tau = 120$ fs, $\vert \mathcal{F} \vert^{2} \approx 5.58 \times 10^{-4}$.


In a previous experimental study of DBT \cite{C.Toninelli2010}, pulsed lasers with an average power of 3 W, a repetition rate of 76 MHz, and resulting peak intensities ranging from 500-4000 kW/cm$^{2}$ were used to excite $GS \rightarrow S_{1}$; for these peak intensities, we predict that the ratio of the populations of the $S_{D_{2}}$ state and the $S_{1}$ state (\ref{eq-ratio-pops}) is on the order of 10$^{-7}$. At a peak intensity of 30 MW/cm$^{2}$ (corresponding to a pulse energy of 0.32 nJ and assuming a circular laser spot with a radius of 50 $\mu$m), our perturbative assumption (i.e. the excitation by the laser pulse is weak) breaks down, and the predicted population of the $S_{1}$ state is large ($\rho_{S_{1}} \approx 0.2$). For this peak intensity of the pulse, the ratio between the populations of the $S_{D_{2}}$ state and the $S_{1}$ state (\ref{eq-ratio-pops}) is on the order of $10^{-6}$. 
This indicates that even for intensities strong enough to significantly populate the $S_{1}$ state, the population of the $S_{D_{2}}$ state will be minuscule relative to the population of the $S_{1}$ state. 
\section{\label{sec:conc} Conclusion}



We applied a method developed for the description of the electronic and optical properties of graphene flakes to study the optical properties of dibenzoterrylene (DBT), a candidate material for single photon source (SPS) applications. We set the Hubbard $U$ parameter of our calculation such that the lowest energy singlet excited state, labeled $S_{1}$, is 1.58 $\,e$V above the ground state energy, in agreement with the experimental value. Our calculated radiative line width for the transition from the ground state to the $S_{1}$ state (denoted $GS \rightarrow S_{1}$) agrees with the experimental value as well. Assuming the experimental measurement of the radiative linewidth is limited only by the excited state lifetime, then our calculated value of the $GS \rightarrow S_{1}$ transition dipole moment is consistent with its experimental value. 

For DBT to be a good SPS, there should be no other optical processes that compete with the transition from the ground state to the first singlet excited state:  There should be no other linear absorption peaks near the peak due to $GS \rightarrow S_{1}$; further absorption from the $S_{1}$ state should not occur at photon energies near the energy of $GS \rightarrow S_{1}$;  inter-system crossing (ISC) to the first triplet state, labeled $T_{1}$, should not be significant; and further absorption from the $T_{1}$ state also should not occur at photon energies near the energy of $GS \rightarrow S_{1}$.


We calculated the oscillator strengths of the bright transitions of DBT up to 6.0 $\,e$V, which should be useful for testing the model against future experiments; excitations from the $sp^{2}$ states, not included in this model, are not expected to be in this energy range. Our calculations predict that there are no other competing linear absorption features near the energy of $GS \rightarrow S_{1}$. We also calculated the further absorption from the $S_{1}$ state; our calculations indicate that there is no further absorption from the $S_{1}$ state for the energy range of interest in the application of DBT  as a SPS. We characterized the charge distributions involved in the bright transitions of DBT, and we showed that the spatial profiles of transitions which have transition dipole moments that are polarized along the $\hat{\mbf{y}}$ axis have electron concentration extended over the entire system, while the spatial profiles of transitions which have transition dipole moments polarized along the $\hat{\mbf{x}}$ axis have electron concentration primarily in the center of the system. 

Our calculations indicate that the $T_{1}$ state has energy 0.85 $\,e$V above the ground state; in our calculation, the energy above the ground state of the $T_{1}$ state is approximately half the energy above the ground state of the $S_{1}$ state.  Such a large difference in energy between the $S_{1}$ state and the $T_{1}$ state indicates that the ISC rate is small in DBT \cite{R.Englman1970,Marian2012}. We calculated the further absorption from the $T_{1}$ state; our calculations indicate that there is no further absorption from the $T_{1}$ state for the energy range of interest in the application of DBT as a SPS.

We also calculated the two-photon absorption (TPA) of DBT. 
An understanding of the nonlinear optical properties of DBT is important for general optical  applications, and also for the specific application of DBT as a SPS since it reveals whether there are any competing nonlinear optical processes against the optical transition that generates the desired photons.
The TPA spectrum showed that in the low photon energy regime, three states are two-photon active: a state composed primarily of HF double excitations, and two states composed mainly of HF single excitations. The strong two-photon absorption occurs at fundamental photon energies near the energy of $GS \rightarrow S_{1}$. If narrow frequency laser pulses (with a temporal full width at half maximum greater than 120 fs)  and weak peak intensities (on the order of 1000 kW/cm$^{2}$), or continuous wave lasers are used to excite $GS \rightarrow S_{1}$, then the strong TPA that occurs at photon energies near the energy of $GS \rightarrow S_{1}$ should not hinder the SPS application of DBT. 

Therefore, our calculations indicate that DBT is a good candidate for a SPS, as there are no other competing absorption features near the energy of the transition from the ground state to the $S_{1}$ state.  The calculations we have performed have also elucidated qualitative features of the higher energy absorption spectrum of DBT, and we expect that these qualitative features will be of interest for considering the use of DBT for other optical applications besides SPS. 

\begin{appendix}

\begin{widetext}
\section{\label{dbt-app}Integrated Third Order Polarizability}
In this appendix, we derive an expression for the integrated third polarizability.
The largest contribution to the third order polarizability \cite{Boyd2008b} is given by
\beq
\alpha^{(3)}_{klop} \left(\omega; \omega, \omega,-\omega \right) \approx \frac{1}{\epsilon_{0} \hbar^{3}} \sum_{vnm} \frac{\mu^{k}_{gv} \mu^{l}_{vn} \mu^{o}_{nm} \mu^{p}_{mg} }{\left(\omega_{vg} - \omega - i\gamma_{vg} \right) \left(\omega_{ng} - 2\omega - i\gamma_{ng} \right) \left(\omega_{mg} - \omega - i\gamma_{mg} \right)}
\label{ap1-eq-3op}
\eeq
near the $\omega_{ng} \approx 2\omega$ resonance. At such frequencies, the third order polarizability (\ref{ap1-eq-3op}) can be further approximated as
\beq
&& \alpha^{(3)}_{klop} \approx  \frac{1}{\epsilon_{0} \hbar^{3}} \sum_{vnm} \frac{\mu^{k}_{gv} \mu^{l}_{vn} \mu^{o}_{nm} \mu^{p}_{mg}}{\left( \omega_{vg} - \overline{\omega} \right) \left( \omega_{mg} - \overline{\omega} \right)} \frac{1}{\left( \omega_{ng} - 2\omega - i\gamma_{ng} \right)}, \nonumber \\
\label{ap1-eq-3op-appr}
\eeq
where $\overline{\omega} = \omega_{ng} / 2$. The imaginary component of (\ref{ap1-eq-3op-appr}) is given by
\beq
&& \text{Im } \left( \alpha^{(3)}_{klop} \right) \approx \frac{1}{\epsilon_{0} \hbar^{3}} \sum_{vnm} \frac{\mu^{k}_{gv} \mu^{l}_{vn} \mu^{o}_{nm} \mu^{p}_{mg}}{\left( \omega_{vg} - \overline{\omega} \right) \left( \omega_{mg} - \overline{\omega} \right)} \frac{\gamma_{ng}}{\left( \omega_{ng} - 2\omega \right)^{2} + \gamma_{ng}^{2}}. \nonumber \\
\label{ap1-eq-3op-appr2}
\eeq
Integrating (\ref{ap1-eq-3op-appr2}) over all frequencies $\omega$, we obtain 
\beq
&& \int^{\infty}_{-\infty} \text{Im } \left( \alpha^{(3)}_{klop} \right) d\omega \approx  \frac{\pi}{2\epsilon_{0} \hbar^{3}} \sum_{vnm} \frac{\mu^{k}_{gv} \mu^{l}_{vn} \mu^{o}_{nm} \mu^{p}_{mg}}{\left( \omega_{vg} - \overline{\omega} \right) \left( \omega_{mg} - \overline{\omega} \right)}, \nonumber \\
\label{ap1-eq-3op-int}
\eeq
where we have used $\int^{\infty}_{-\infty} 1 / \left( \left( 2x-x_{0} \right)^2 + \gamma_{ng}^{2} \right)  dx = \pi / 2\gamma_{ng}$. The expression (\ref{ap1-eq-3op-int}) is the integrated third order polarizability, which is independent of the frequency broadening $\gamma$. Therefore, the strength of the two photon transition $GS \rightarrow Z$ is given by
\beq
B_{klop} \left( GS \rightarrow Z \right) =  \frac{\pi}{2\epsilon_{0} \hbar^{3}}\sum_{vm} \frac{ \mu^{k}_{gv} \mu^{l}_{vZ} \mu^{o}_{Zm} \mu^{p}_{mg}}{\left( \omega_{vg} - \overline{\omega} \right) \left( \omega_{mg} - \overline{\omega} \right)},
\label{ap1-eq-3op-str}
\eeq
which is Eq.~\eqref{eq:Bklop}.
\end{widetext}
\end{appendix}

\bibliographystyle{unsrt}
\bibliographystyle{apsrev4-1}
\bibliography{dbt_pap_optics_only_bib}

\end{document}